\documentclass[11pt,twoside]{article}

\usepackage{asp2006}
\usepackage{epsf}
\usepackage{lscape}
\usepackage{graphicx}

\begin{document}
\title{Star Formation Rate Determinations}   
\author{Daniela Calzetti}   
\affil{Dept. of Astronomy, University of Massachusetts, Amherst, MA, U.S.A.}    

\begin{abstract} 
I  review determinations of star formation rate (SFR) indicators from the ultraviolet to 
the infrared, in the context of their use for galaxies and 
galaxy surveys. The mid-infrared SFR indicators have garnered
interest in recent years, thanks to the Spitzer capabilities and the opportunities 
offered by the upcoming Herschel Space Telescope. I discuss what we have learned in the 
mid-infrared from studies of local galaxies combining Spitzer with HST and other data.
\end{abstract}


\section{Introduction}

Star formation is one of the fundamental processes driving the evolution of galaxies. It depletes 
galaxies of gas (and dust) by converting it to stars, controls the metal enrichment of  the 
interstellar  (ISM) and intergalactic medium (IGM), regulates the radiative and mechanical 
feedback into the ISM and IGM, and shapes the stellar populations' mix. Star 
formation is the true link between the invisible Universe \citep[driven by gravity, and traced by 
models of galaxy formation and evolution, e.g., ][]{kat92,spr03,spr05,sti06} and the visible Universe that 
is typically measured, one example being the redshift evolution of the 
cosmic star formation rate density in comoving space 
\citep[e.g.,][]{mad96,lil96,sma97,mad98,hug98,ste99,eale99,yan99,bar00,wil02,gia04,nor04,cha05,hop06}. Thus, it 
becomes paramount to characterize and physically quantify the laws of star formation 
\citep{ken98a,ken07}, and to obtain accurate measures of the star formation rates of galaxies.

Star formation rates indicators have been defined at, basically, all wavelengths across 
the spectrum, from the X--ray to the radio 
\citep[see, e.g.,][and references therein]{ken98b,kew02,ran03,bel03,hop04,kew04,cal05,sch06,cal07}. 
The common characteristic of all indicators is that they all probe the {\em massive stars formation 
rate}. To convert this measure to a SFR, an assumption on the stellar Initial Mass Function (IMF) 
needs to be made; this problem is being covered in other presentations at this conference, together 
with the stochastic effects present at low star formation levels (see, e.g., P. Kroupa's contribution to 
this conference).  

This review mostly concentrates on the robustness of converting the luminosity L($\lambda$) of a whole 
galaxy  into a SFR indicator, and the potential impact of  
contributions to L($\lambda$) that are not strictly linked to the current star formation. Examples of such 
`extraneous' contributions are evolved stellar populations, which may contribute to 
the light emitted at a given wavelength but are not part of the star--forming population, or  the 
presence of dust. 

For sake of brevity, I will only touch upon SFR indicators at ultraviolet (UV), optical, and 
infrared (FIR) wavelengths. 
Indicators at UV and optical/near--infrared indicators only probe the stellar light that emerges from 
galaxies {\em unabsorbed by dust}. Infrared SFR indicators are complementary to UV--optical 
indicators, because they measure star formation via the stellar light that has been reprocessed by 
dust and emerges beyond a few $\mu$m. 


\section{The Ultraviolet}

The ultraviolet ($\lambda\sim$912--3000~\AA) emission from galaxies is extensively used 
as a SFR indicator, especially in the 
medium/high redshift regime, where the restframe UV moves into the optical observer's frame. Recently, 
GALEX \citep{mar05}, with its UV survey capability, has revamped the interest of this wavelength 
regime for studies in the local Universe \citep[e.g.,][]{sal07}. 

\begin{figure}[!ht]
\plottwo{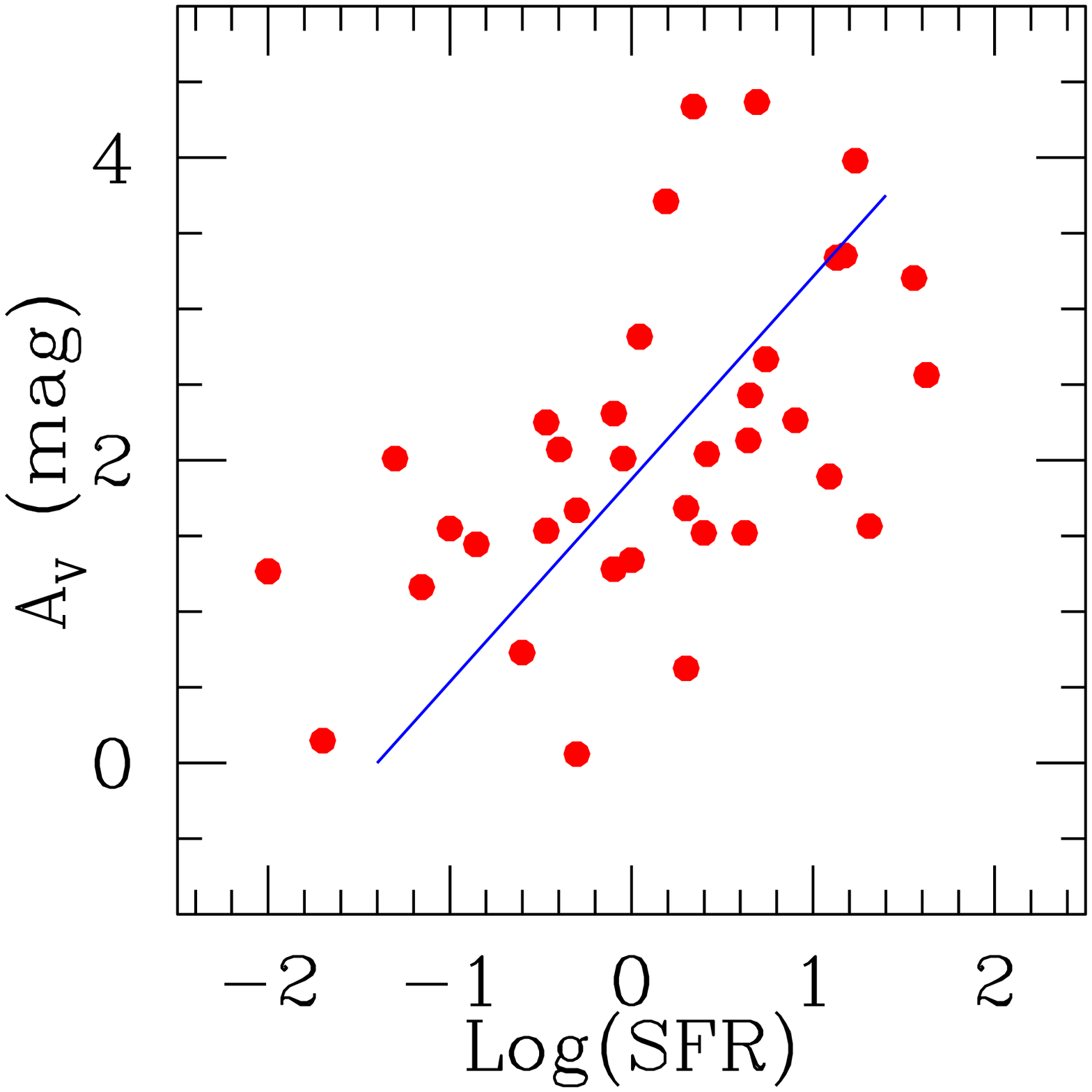}{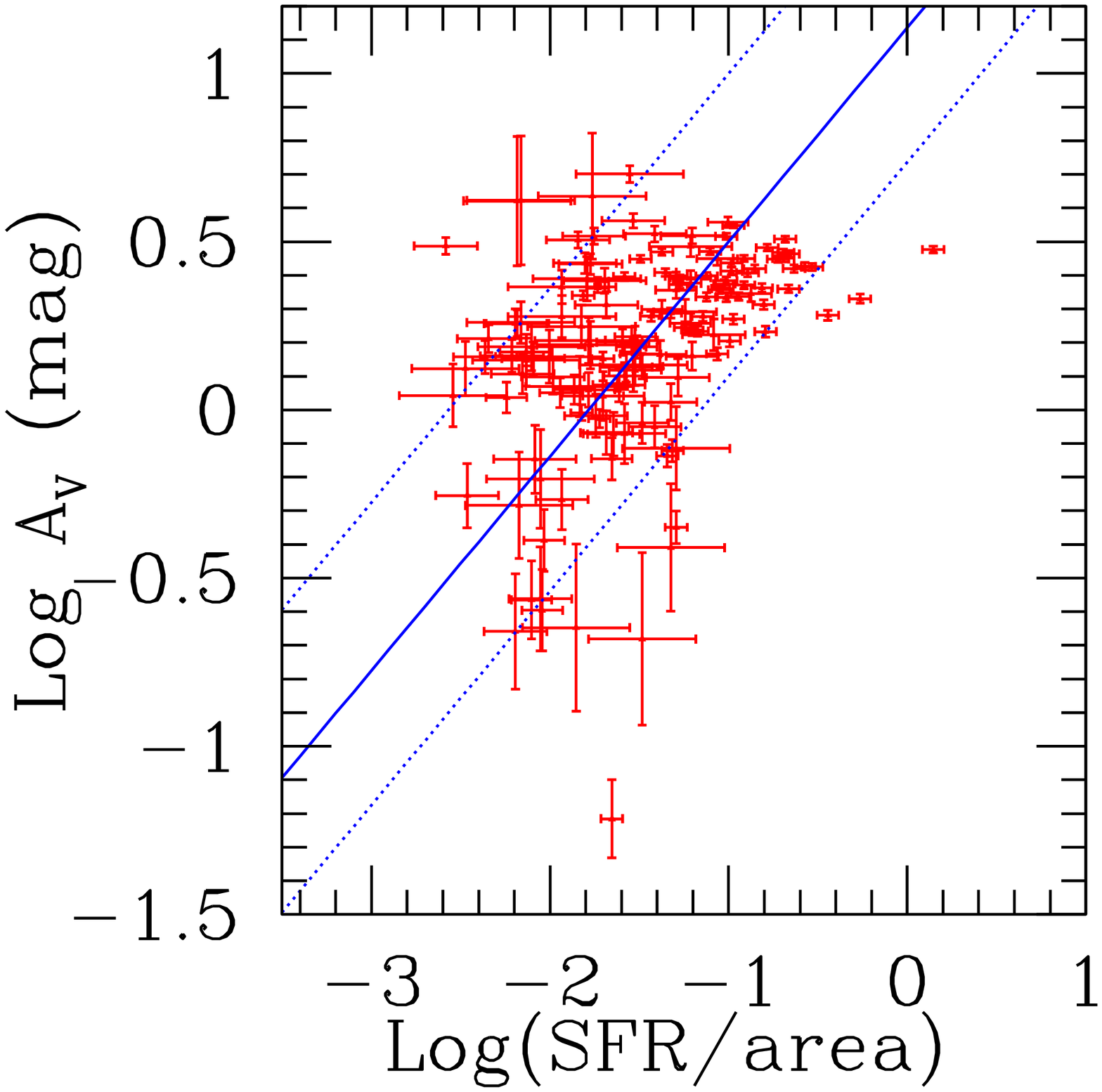}
\caption{{\bf Left.} The attenuation A$_V$ (in magnitudes), derived from hydrogen nebular line emission 
\citep[H$\alpha$/H$\beta$ or H$\beta$/Br$\gamma$,][]{cal96} as a function of the log(SFR) 
(in units of M$_{\odot}$~yr$^{-1}$) for a sample of local starbursts; the continuous line shows the 
mean trend of the datapoints. {\bf Right.} As similar plot to the one at left (but in log--log scale), this 
time for 164 star--forming regions in 21 nearby galaxies. All the galaxies have metallicity close to 
the solar value. A$_V$ (from the H$\alpha$/P$\alpha$ ratio) is in units of magnitudes, and the 
SFR density in the horizontal axis is in units of   M$_{\odot}$~yr$^{-1}$~kpc$^{-2}$. The continuous 
line is the expected trend after 
combining the Schmidt--Kennicutt law with the dust--to--gas ratio of the galaxies 
\citep{cal07}. The dotted lines are the 90\% boundary to the datapoints.\label{fig1}}
\end{figure}

The advantage of using the UV is that it directly probes the bulk of the emission 
from young, massive stars. Its main disavantage, however, is its high sensitivity to dust reddening 
and attenuation. 
For reference, a star--forming galaxy with A$_V$=1~mag has A$_{1500~\AA}\approx$3~mag, but the 
specific value will depend on both the dust amount {\em and} the dust geometry in the galaxy. 
To make matters worse, there is a well--established correlation between attenuation and SFR
 in galaxies and star--forming regions 
\citep[Figure~\ref{fig1},][]{wan96,heck98,cal01,hop01,sul01,bua02,cal07}; thus, 
the galaxies that are most actively forming stars and are, therefore, likely to be important 
in the cosmic SFR census, are also those whose UV luminosity requires the largest corrections 
for the effects of dust attenuation.  An increase of a factor 100 in SFR implies on average an 
increase in the attenuation by A$_V\approx$1--3~mag, for star--forming galaxies. The quantitative 
details may depends on the exact nature of the star--forming system, but the qualitative trend 
is common to most (Figure~\ref{fig1}). At fixed metallicity \citep[and dust--to--gas ratio,][]{dra07}, 
this trend is a consequence of the correlation between SFR density and gas density 
\citep{ken98a,ken07}. A number of authors \citep[e.g.,][]{cal94,meu99,cal00,hop01,sul01,bua02,bua05,bel03,hop04,igl06,sal07} have 
derived techniques to correct the observed UV emission from galaxies for the effects of dust 
attenuation. The effectiveness of these techniques, however, depends in general on the nature of the 
galaxy population they are being applied to. 

The additional complication with using the UV as a SFR indicator is the long timescale (of--order 100~Myr) in which stars are relatively luminous in the non--ionizing UV wavelength range.  
The cross--calibration of SFR(UV) with indicators at other wavelengths may be complicated by 
the different timescales involved. This may be the case when comparing SFR(UV) with, e.g., 
SFR(H$\alpha$) (see next section), where the latter has a timescale about 10 times shorter 
than the former. In this case, the star formation history of the system plays a significant role. 
 If an unresolved object is only measured, say, at 1500~\AA, it will be impossible to know whether 
 the emission is coming from a system that has been forming stars at the constant level of 
 1~M$_{\odot}$~yr$^{-1}$ for the past 100~Myr, 
 or a system that is no longer forming stars but has been passively evolving for 
 the past 50~Myr after a 4$\times$10$^8$~M$_{\odot}$ burst of star formation \citep{lei99}.  

\begin{figure}[!ht]
\plotone{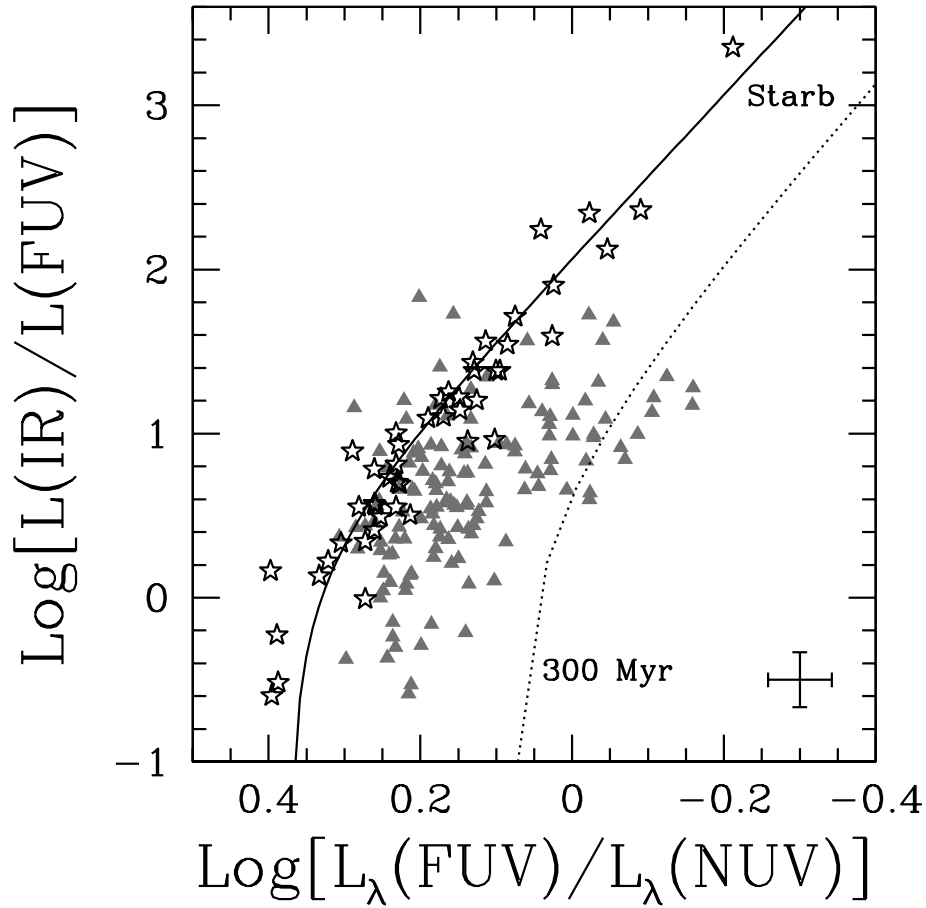}
\caption{The FIR/UV ratio (the ratio of the far--infrared to the far--UV luminosity) versus 
the UV~color (given here as the ratio between the GALEX far--UV and near--UV 
fluxes) for starburst galaxies and quiescently star--forming regions. The starburst galaxies are 
shown with star symbols. The grey filled triangles are star--forming regions within the 
galaxy NGC5194 \citep{cal05}. Redder UV colors correspond on average to larger FIR/UV 
ratios. The continuous line shows the best fit to the starburst galaxies, 
which is well represented by a model of a progressively more attenuated (from left to right) 
constant star--forming population. The dotted line shows the same dust attenuation trend 
for a 300~Myr old stellar population; this model represents a reasonable lower envelope 
to the NGC5194 star--forming regions.\label{fig2}}
\end{figure}

The $\approx$10 times longer stellar timescales probed by the UV relative to tracers of 
ionizing photons accounts for the difference in the properties of the dust attenuation between 
starburst  and quiescent star--forming systems. Starburst galaxies show 
a well--defined correlation in the FIR/UV--versus--UV~color plane \citep[Figure~\ref{fig2}][]{meu99}. 
The FIR/UV ratio is a measure of the UV attenuation suffered by the system: the higher the dust 
attenuation, the larger the amount of UV stellar energy reprocessed into the FIR by dust. Variations in 
the  UV~color probe the amount of dust reddening present in the system. Thus, in starbursts, 
 the UV reddening is a good tracer of the total UV dust attenuation \citep{cal01}. 
 Quiescent star--forming galaxies and regions 
show a $\sim$10~times larger spread in the FIR/UV ratio than starburst galaxies, 
at fixed UV~color; the spread is in  the sense that the starburst galaxies form the upper 
envelope to the quiescently star--forming systems \citep[Figure~\ref{fig2} and][]{bua02,bua05,bel02,gor04,kon04,sei05,cal05}.  Thus, a far larger range of UV dust attenuations 
seems to be present in star--forming objects than in starbursts, for the same UV~color. This 
is possible if in quiescent star--forming systems the reddening of the UV~colors is not only a probe 
of dust, but also of ageing of the individual star--forming regions  contributing to the 
UV emission in the system \citep[regions up to an age of $\approx$100--300~Myr,][]{cal05}. Specifically, 
older clusters can contribute significantly to the measured UV emission in these galaxies, as they 
are not as attenuated by dust as younger systems. This will complicate the definition of SFR(UV) in 
quiescent star--forming galaxies, and change its calibration relative to the starburst galaxies 
\citep[e.g.,][]{sei05,cor06,sal07}.

\section{The Optical and Near--Infrared}

SFR indicators in the optical and near--infrared wavelength range ($\lambda\sim$3000--25,000~\AA) 
are derived from the many hydrogen recombination lines (mainly H$\alpha$, H$\beta$, P$\beta$, 
P$\alpha$, Br$\gamma$) and from forbidden line emission (chiefly [OII] and [OIII]). These 
indicators, thus, trace the ionizing photons in the system. The short lifespan of massive, 
ionizing stars confines the timescale probed by the optical indicators to about 10~Myr, 
implying that they are tracers of the {\em current} SFR. 

Calibrations for these indicators, in particular for H$\alpha$($\lambda$6563~\AA) and 
for [OII]($\lambda$3727~\AA), have been presented by many authors 
\citep[to name a few,][and references therein]{ken98b,gal89,jan01,ros02,kew02,cha01,kew04,mou06}. 

There are a number of effects that impact the SFRs derived from optical and near--infrared 
line emission. First and foremost, dust extinction, which affects the blue lines more than the 
red ones. For example, if extinction corrections are neglected, the median SFR derived for 
a generic sample of nearby galaxies will be underestimated by roughly a factor of 3 if using H$\alpha$, and by 
about twice as much if using [OII] \citep{ros02}. These mean values do not convey 
the fact that brighter galaxies are also more extincted, as discussed in the previous section. 
In starburst galaxies, dust attenuation is about a factor of two larger 
for the ionized gas than for the stellar continuum, at the same wavelength \citep{cal94}: the 
dust attenuation in the H$\beta$($\lambda$4861~\AA) line is the same that in the stellar  continuum 
at $\sim$2300~\AA. 

Line emission SFR indicators are also sensitive to the upper end of the stellar IMF, i.e., the 
number of ionizing stars formed, much more than the UV stellar continuum. Changing the upper 
stellar mass of the IMF from 100~M$_{\odot}$ to 30~M$_{\odot}$ decreases the number of 
ionizing photons by a factor of $\sim$5, and the UV stellar continuum at 1500~\AA~ by a factor 2.3 
\citep[from the models of][]{lei99}. Other factors to take into account for the line emission SFR 
indicators are the underlying stellar absorption in the case of the hydrogen recombination lines 
\citep{ros02}, and the metallicity and ionization conditions for the forbidden lines \citep[e.g., ][]{mou06}.  

For reference, for a solar metallicity stellar population with constant SFR over the past 100~Myr, 
forming stars with an IMF  consisting of two power laws, with slope $-$1.3 in the range 
0.1--0.5~M$_{\odot}$ and slope $-$2.3 in the range 0.5--120~M$_{\odot}$, the SFR(H$\alpha$) 
is given by: 
\begin{equation}
SFR(M_{\odot}~yr^{-1}) = 5.3 \times 10^{-42} L(H\alpha) (erg~s^{-1}).
\end{equation}
Variations of $\pm$20\% on the calibration in this relation are present for younger/older 
ages and metallicities down to $\sim$1/5th solar. The $\sim$50\% difference between 
the calibration in Equation~1 and that of  \citet{ken98b} is mainly due to differences in 
assumptions on the stellar IMF and the age of the stellar populations  (100~Myr in our case versus 
infinite age in \citet{ken98b}).  

\section{The Infrared}

The earliest calibrations of the infrared ($\lambda\sim$5--1000~$\mu$m) dust emission as 
a SFR indicator date back to IRAS \citep[see, for a review and an independent calibration,][]{ken98b}. 
The connection between star formation and infrared emission is based on the fact that, at least 
in the nearby Universe, star--forming regions tend to be dusty and the dust absorption 
cross--section peaks in the UV (also the peak of emission of young, massive stars). 

Since the very beginning, however, it was clear that the calibration of the FIR emission as a 
SFR tracer carried a number of issues \citep{hun86,per87,row89,dev90,sau92}. First, 
not all of the luminous energy produced by recently formed stars is
re-processed by dust in the infrared; in this case, the FIR only recovers part of the SFR, and 
the fraction recovered depends, at least partially, on the amount of dust in the system. 
Second, evolved, i.e., non-star-forming, stellar populations also heat the dust that emits in the 
FIR wavelength region, thus affecting the calibration of SFR(FIR) in a
stellar-population-dependent manner. Third,  SFR(FIR) is a `calorimetric' measure (the full energy 
census in the range $\sim$5--1000~$\mu$m needs to be included), and the FIR 
spectral energy distribution (SED) depends on the stellar field intensity \citep{hel86,drli07}; thus, 
extrapolations of SFR(FIR) from too a sparsely sampled SED are subject to large uncertainties. 

The use of monochromatic  (i.e., one band or wavelength) infrared  emission 
for measuring SFRs  offers one definite  advantage over the bolometric infrared luminosity:  it 
removes the need for  uncertain extrapolations of the dust SED 
across the full wavelength range. The interest in calibrating monochromatic mid-infrared SFR diagnostics stems from their potential application to both the 
local Universe \citep[e.g., in the investigation of the scaling laws of star formation][]{ken07} and intermediate and high redshift galaxies observed with Spitzer, Herschel and
future millimiter/radio facilities \citep[e.g.][]{daddi05,wu05}. 

Studies using ISO data provided the first window on the use of the monochromatic mid--IR 
(MIR, $\lambda\sim$5--40~$\mu$m) emission as a SFR indicator; investigations with 
the higher angular resolution and more sensitive Spitzer 8~$\mu$m and 24~$\mu$m data 
have expanded on the ISO results. The dust emission in the 
MIR wavelength range is characterized by both continuum and bands. The continuum is 
due to dust heated by a combination of single--photon and thermal equilibrium processes, with 
the latter becoming more and more prevalent over the former at longer wavelengths \citep{drli07}. 
The interest in calibrating the MIR dust continuum as a SFR indicator stems from the consideration 
that the dust heated by hot, massive stars can have high temperatures and will then emit at 
short infrared wavelengths. 
 The MIR bands are generally attributed to Polycyclic Aromatic
Hydrocarbons \citep[PAH, ][]{lege84,sell84}, large molecules transiently heated
by single UV and optical photons in the general radiation field of galaxies or near B stars
 \citep{li02,peete04,matt05}, and which can be
destroyed, fragmented, or ionized by harsh UV photon fields \citep{boul88,pety05}. 

 \begin{figure}[!ht]
\plottwo{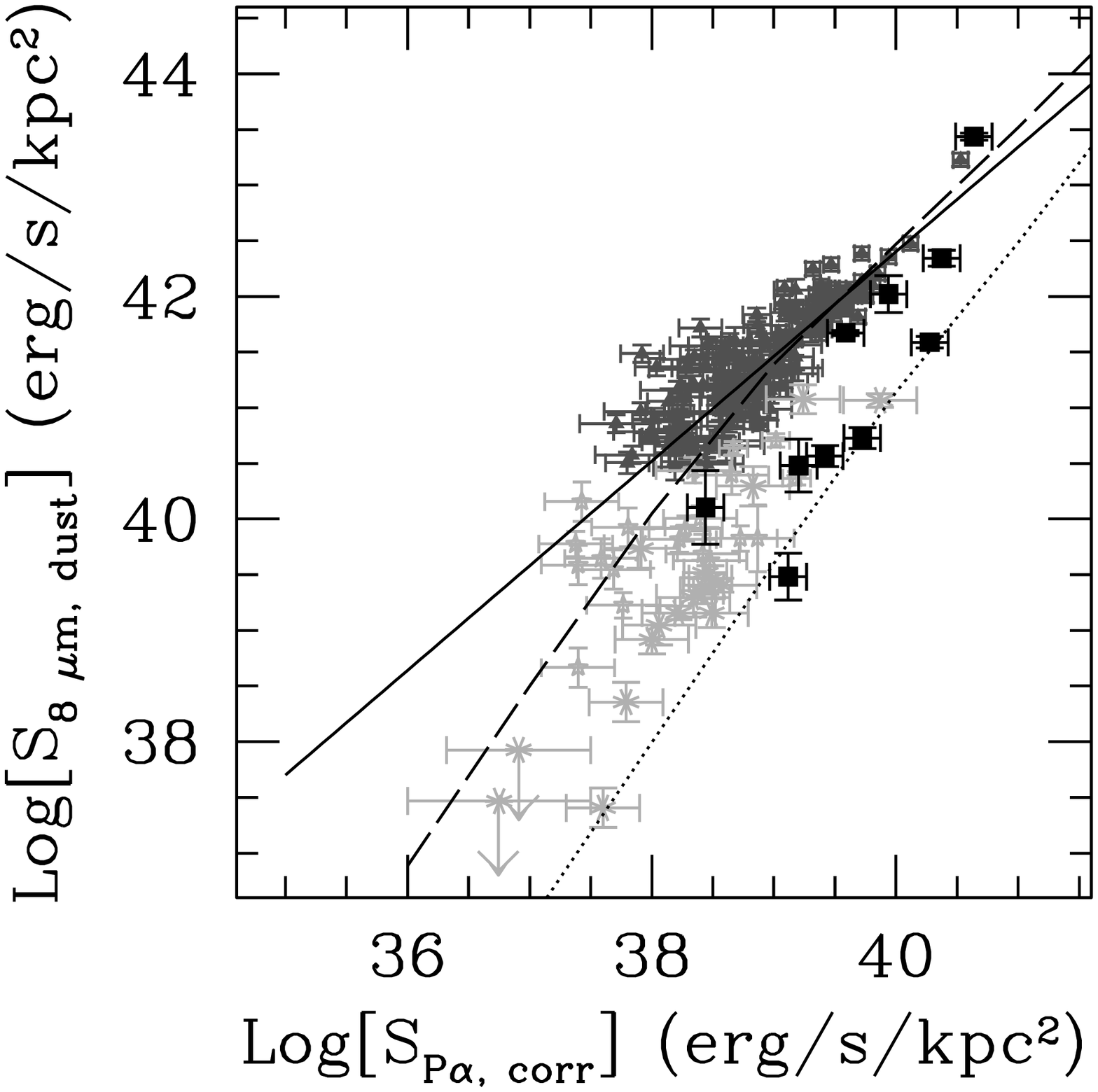}{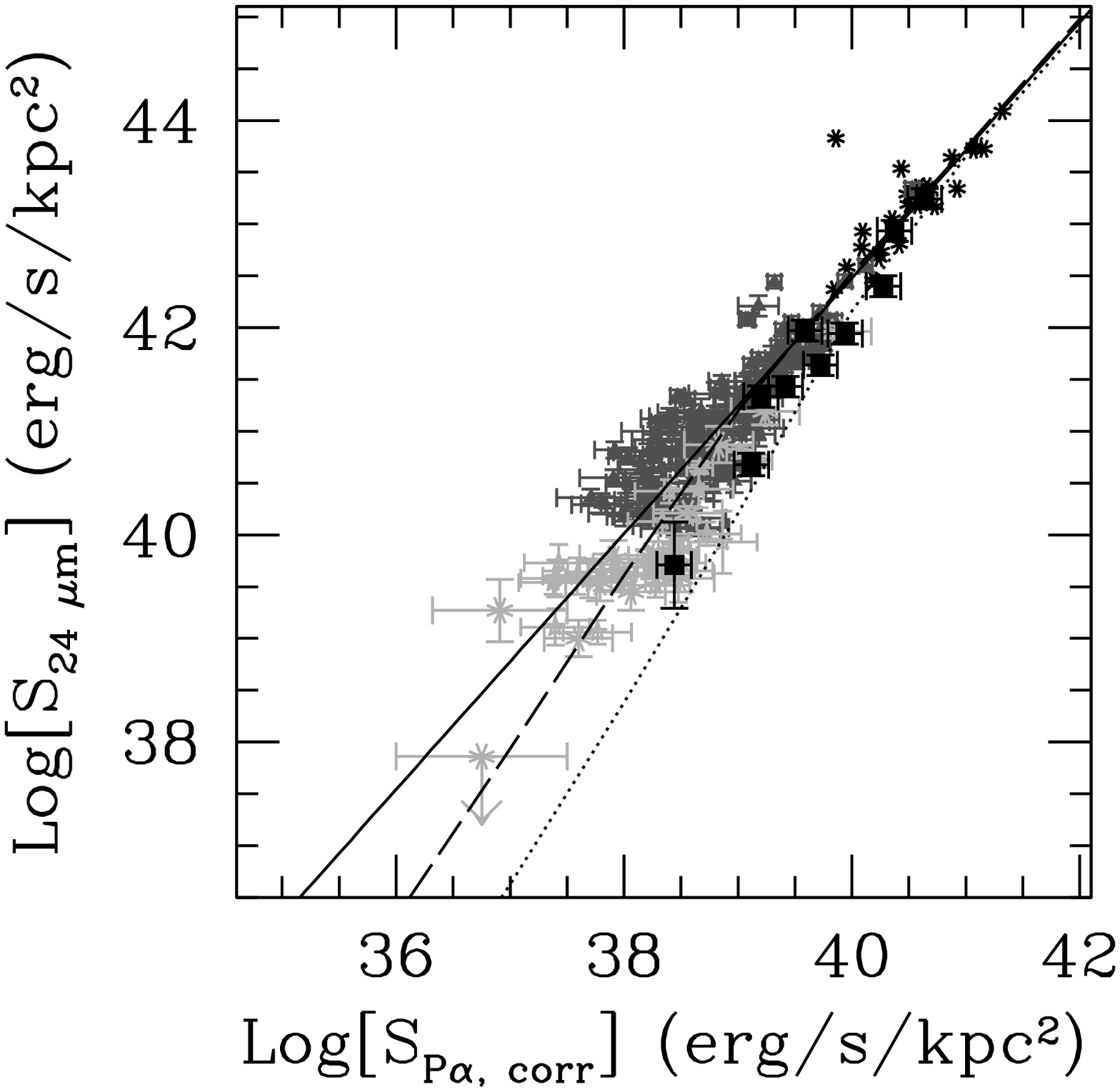}
\caption{{\bf Left.} The luminosity surface density (LSD=luminosity/area) at 8~$\mu$m versus the 
P$\alpha$ LSD of 10 starburst galaxies  and 220 star--forming regions in 33 nearby galaxies
\citep{cal07}. The 8~$\mu$m emission is from Spitzer images, and has the stellar continuum subtracted. 
The P$\alpha$ line emission ($\lambda$=1.876~$\mu$m) is from HST/NICMOS images, 
has been corrected for dust extinction using the H$\alpha$/P$\alpha$ ratio, and is used here as an 
unbiased tracer of massive stars SFR. Of the 220 regions, the $\sim$180 regions in 
high--metallicity galaxies, 12$+$log(O/H)$>$8.3, are marked in dark grey, and the $\sim$40 regions in 
low--metallicity galaxies are  in light grey. The starburst galaxies \citep{enge05} are the black squares. 
The continuous line is the best fit  to the high--metallicity star--forming regions (dark grey), and has 
slope 0.94. Models for a young stellar population with increasing amount of star formation and dust 
are shown as a dash line (Z=Z$_{\odot}$) and a dot line (Z=1/10~Z$_{\odot}$), using the stellar 
population models of \citet{lei99} and the dust models of \citet{drli07}. The spread in the datapoint 
around the best fit line is well accounted for by a spread in the stellar population's age in the range 
2--8~Myr. {\bf Right} The same as a left panel, with the vertical axis now reporting the 24~$\mu$m 
LSD, also from Spitzer. The black asterisks are the Luminous Infrared Galaxies (LIRGS) from \citet{alon06}. The best fit line (continuous line) has slope 1.2.\label{fig3}}
\end{figure}

 \citet{rous01} and \citet{forst04} showed that the emission 
in the 6.75~$\mu$m ISO band correlates with the number of ionizing photons in galaxy disks 
and in the nuclear regions of galaxies. Conversely, \citet{bose04} found that the mid--IR emission  
in a more diverse sample of galaxies (types Sa through Im--BCDs) correlates more closely 
with tracers of evolved stellar populations not linked to the current star formation.  Additionally,
\citet{haas02} found that the ISO 7.7~$\mu$m emission is correlated with the 
850~$\mu$m emission from galaxies, suggesting a close relation between the ISO band 
emission and the cold dust heated by the general (non--star--forming) stellar population. 
Spitzer data of the nearby galaxies NGC300 and NGC4631 show that the 8~$\mu$m band 
emission highlights the rims of
HII regions and is depressed inside the regions, indicating that the PAH dust is heated in the 
photo--dissociation regions  surrounding HII regions and is destroyed within the regions 
\citep{helo04,bend06}. Analysis of the mid--IR emission from the First Look 
Survey galaxies shows that the correlation between the 8~$\mu$m band emission 
and tracers of the ionizing photons is shallower than unity \citep{wu05}, in agreement 
with the correlations observed for HII regions in the nearby star--forming galaxy NGC5194 
 \citep{cal05}. 
 
A recent analysis of the 8~$\mu$m emission from 220 star--forming regions in 33 nearby galaxies 
from the project SINGS \citep[the Spitzer Infrared Nearby Galaxies Survey][]{ken03} has shown 
that this MIR band is sensitive to both metallicity and star formation history
\citep[Figure~\ref{fig3} left, and][]{cal07}.  The dependence on metallicity 
\citep[e.g.,][]{bose04,enge05,dra07} implies that regions with values about 1/10~Z$_{\odot}$ 
are about 30 times underluminous at 8~$\mu$m relative to regions with solar metallicity and same 
SFR. The dependence on star formation history is due to the contribution to the 8~$\mu$m emission 
from dust heated by non--ionizing stellar populations that are unrelated to the current star formation 
\citep{peete04,cal07}. However, for a restricted choice of parameters, i.e., actively 
star--forming regions with similar (e.g., solar) metallicity, the 8~$\mu$m emission correlates almost 
linearly with the SFR (Figure~\ref{fig3}, left). 

The same analysis shows that the 24~$\mu$m emission is, conversely, a good SFR tracer for 
galaxies,  in the absence of strong AGNs \citep[Figure~\ref{fig3}, right, and][]{cal07}. Similar 
conclusions had been reached by \citet{cal05} and by \citet{gonz06} for star--forming regions 
in NGC5194 and NGC3031, respectively, and by \citet{wu05} and \citet{alon06} for bright  
galaxies. In particular, \citet{cal05} has concluded that the 24~$\mu$m emission is more 
closely related to the H$\alpha$ (ionizing photons) emission than to the UV (non--ionizing stellar 
continuum) emission. A single correlation can be defined over 3.5 orders of magnitude in 
luminosity surface density (luminosity/area), yielding 
the following SFR calibration:
\begin{equation}
SFR(M_{\odot}~yr^{-1}) = 1.24\times 10^{-38} [L(24~\mu m) \ (erg~s^{-1})]^{0.88} .
\end{equation}
This calibration is very similar to that of \citet{alon06}, who have 
used a sample of Ultraluminous Infrared 
Galaxies, LIRGs, and NGC5194. The non--linear trend of L(24~$\mu$m) with SFR is a direct 
consequence of the increasing dust temperature for more actively star forming 
objects \citep{drli07}; higher dust temperatures correspond to higher fractions 
of the dust emission emerging in the MIR. 

The 24~$\mu$m emission has a much lower sensitivity to metallicity than the 8~$\mu$m emission, 
decreasing by just a factor 2-4 for a tenfold decrease in metallicity. This small decrease if fully 
accounted by the  increased transparency (lower dust--to--gas ratio) of the medium for lower 
metal abundances \citep{dra07}. In addition, there seem to be very little contribution to the 24~$\mu$m 
emission from non--ionizing stellar populations, at least within the limits of the \citet{cal07} analysis. 

\begin{figure}[!ht]
\plotone{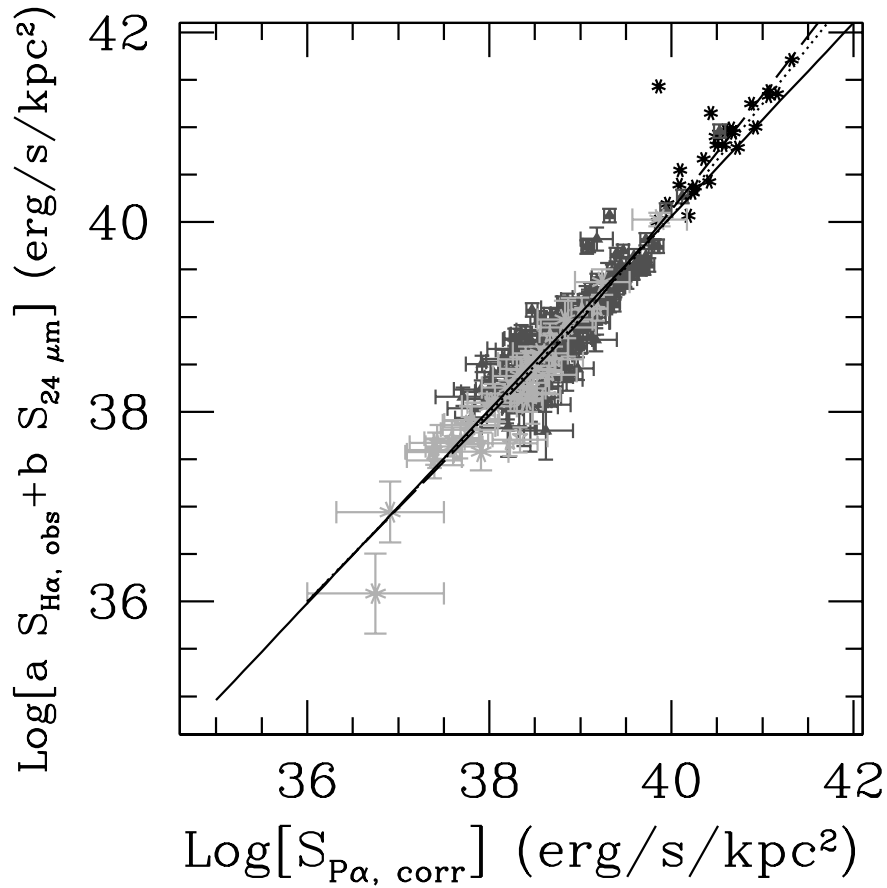}
\caption{As in Figure~\ref{fig3}, for the combined H$\alpha$ and 24~$\mu$m LSD. 
For this SFR indicator, data and models are degenerate in metallicity. The slope of 
the best fit line is unity, and the data follow the 1--to--1 line up to the LIRGs. 
For the brightest galaxies, however, there are deviations from this trend, 
reflecting the increasing temperature of the dust in a regime where the 24~$\mu$m emission 
(i.e., the dust--obscured SFR) dominates over the H$\alpha$ emission (the unobscured SFR). 
\label{fig4}}
\end{figure}

The sensitivity of the SFR indicator to variations in the dust content of the system can be removed 
by combining two tracers: one that probes the dust--obscured star formation and one 
that probes the unobscured one. Combining the 24~$\mu$m luminosity  with 
the {\em observed} H$\alpha$ luminosity (Figure~\ref{fig4}) yields a new SFR indicator, with calibration
 \citep{ken07,cal07}:
\begin{equation}
SFR(M_{\odot}~yr^{-1}) = 5.3\times 10^{-42} [L(H\alpha)_{obs} + (0.031\pm0.006) L(24~\mu m)],
\end{equation}
from equation~1. The calibration in equation~3 is unaffected not only by metallicity variations 
(Figure~\ref{fig4}), but also by stellar population mix. However, it shows some deviation 
from a simple, linear correlation at high luminosity surface densities \citep[Figure~\ref{fig4}, and][]{cal07}, which simply reflects the non--linear trend of L(24~$\mu$m) with SFR once luminosities are 
sufficiently high that the infrared emission dominates over the line emission. 

\section{Summary}

The SFRs based on the UV stellar continuum and the optical/near--infrared line emission probe 
the stellar light that emerges from a galaxy unabsorbed by dust. They both require corrections for 
the effects of dust, which can change the mean values of local samples by factors of few.  In addition, 
comparisons of calibrations between SFR(UV) and SFR(line) require care, as the former probes 
stellar timescales that are at least an order of magnitude longer than the latter. 

In the infrared, the calorimetric SFR(FIR) is based on measuring the stellar energy re--processed by 
dust and suffers from difficult--to--quantify contributions from 
non--star--forming stellar populations. Often SFR(FIR) needs to be extrapolated from sparse 
wavelength sampling, due to observational limitations, which leads to uncertainties.

Monochromatic mid--infrared SFR indicators offer a valid complement to the stellar--based 
SFR tracers and to SFR(FIR). SFR(MIR) tend to be more closely associated to the regions of 
ionizing photon emission than to regions of UV emission. In the absence of AGNs, L(24~$\mu$m) 
and a linear combination of L(H$\alpha$) and L(24~$\mu$m) provide relatively robust SFR 
tracers, with uncertainties of--order a factor 2. Conversely, the 8~$\mu$m emission  is strongly 
dependent on both variations in metallicity (factor of 30 variation in luminosity for a factor $\sim$10 
variation in metallicity) and in stellar population mix (factor of a few variation in luminosity). Equations~2 and 3 should be generally applicable to regions and galaxies dominated by current star formation. 

\acknowledgements 

This work would not have been possible without the combined efforts of the SINGS (Spitzer 
Infrared Nearby Galaxies Survey) team, whom the author thanks heart-fully. Special thanks 
are due to the SINGS project's Principal Investigator, Robert Kennicutt, and collaborators Chad 
Engelbracht, Bruce Draine, and Fabian Walter.  

The author is grateful to Johan Knapen for all his excellent efforts in organizing this 
 stimulating conference, and to John Beckman for offering the perfect excuse for it. 



\begin{thebibliography}{}
\bibitem[Alonso--Herrero et al.(2006)]{alon06} Alonso--Herrero, A., Rieke, G.H., Rieke, M.J., Colina, 
L., Perez-Gonzalez, P.G., \& Ryder, S.D. 2006, \apj, 650, 835
\bibitem[Barger, Cowie \& Richards(2000)]{bar00} Barger, A.J., Cowie, L.L., \& Richards, E.A. 2000, 
\aj, 119, 2092
\bibitem[Bell(2003)]{bel03} Bell, E.F. 2003, \apj, 586, 794
\bibitem[Bell et al.(2002)]{bel02} Bell, E.F., Gordon, K.D., Kennicutt, 
R.C., \& Zaritsky, D. 2002, \apj, 565, 994
\bibitem[Bendo et al.(2006)]{bend06} Bendo, G.J., Dale, D.A., Draine, B.T., Engelbracht, C.W., 
Kennicutt, R.C., Calzetti, D., Gordon, K.D., Helou, G., Hollenbach, D., Li, AIgen, Murphy, E.J., et al. 
2006, \apj, 652, 283
\bibitem[Boselli, Lequeux \& Gavazzi(2004)]{bose04} Boselli, A., Lequeux, J.,
  \& Gavazzi, G. 2004, \aap, 428, 409
  \bibitem[Boulanger et al.(1988)]{boul88} Boulanger, F., Beichmann, C.,
Desert, F.--X., Helou, G., Perault, M., \& Ryter, C. 1988, \apj, 332,
328
\bibitem[Buat et al.(2002)]{bua02} Buat, V., Boselli, A., Gavazzi, G., \&
  Bonfanti, C. 2002, \aap, 383, 801
\bibitem[Buat et al.(2005)]{bua05} Buat, V., Iglesias-Paramo, J., Seiber, M., Burgarella, D., 
Charlot, S., Martin, D.C., Xu, C.K., Heckman, T.M.,, Boissier, S., Boselli, A., et al.  2005, \apj, 619, L51 
\bibitem[Calzetti(2001)]{cal01} Calzetti, D. 2001, \pasp, 113, 1449
\bibitem[Calzetti et al.(2000)]{cal00} Calzetti, D., Armus, L.,
Bohlin, R.C., Kinney, A.L., Koornneef, J., \& Storchi-Bergmann,
T. 2000, \apj, 533, 682
\bibitem[Calzetti et al.(2005)]{cal05} Calzetti, D., Kennicutt, R.C.,
Bianchi, L., Thilker, D.A., Dale, D.A., Engelbracht, C.W., Leitherer, C., 
Meyer, M.J., et al. 2005, \apj, 633, 871
\bibitem[Calzetti et al.(2007)]{cal07} Calzetti, D., Kennicutt, R.C., Engelbracht, C.W., Leitherer, C., 
Draine, B.T., Kewley, L., Moustakas, J., Sosey, M., et al. 2007, \apj, in press (astroph/0705.3377)
\bibitem[Calzetti, Kinney \& Storchi--Bergmann(1994)]{cal94} Calzetti, 
D., Kinney, A.L., \& Storchi--Bergmann, T. 1994, \apj, 429, 582
\bibitem[Calzetti, Kinney \& Storchi--Bergmann(1996)]{cal96} Calzetti, 
D., Kinney, A.L., \& Storchi--Bergmann, T. 1996, \apj, 458, 132
\bibitem[Chapman et al.(2005)]{cha05} Chapman, S.C., Blain, A.W., Smail, I., \& Ivison, R.J. 2005, 
\apj, 622, 772
\bibitem[Charlot \& Longhetti(2001)]{cha01} Charlot, S., \& Longhetti, M. 2001, \mnras, 323, 887
\bibitem[Cortese et al.(2006)]{cor06} Cortese, L., Boselli, A., Buat, V., Gavazzi, G., Boissier, S., 
Gil de Paz, A., Seibert, M., Madore, B.F., \& Martin, D.C. 2006, \apj, 637, 242
\bibitem[Daddi et al.(2005)]{daddi05} Daddi, E., Dickinson, M., Chary,
R., Pope, A., Morrison, G., Alexander, D.M., Bauer, F.E., Brandt,
W.N., Giavalisco, M., Ferguson, H., Lee, K.-S., Lehmer, B.D.,
Papovich, C., \& Renzini, A. 2005, \apj, 631, L13
\bibitem[Desert, Boulanger \& Puget(1990)]{dese90} Desert, F.--X., Boulanger, F., \& Puget, J.L. 1990, \aap, 237, 215
\bibitem[Devereux \& Young(1990)]{dev90}  Devereux, N.A., \& Young, J.S. 1990, \apj, 350, L25
\bibitem[Draine \& Li(2006)]{drli07} Draine, B.T., \& Li A. 2007, ApJ, 657, 810. 
\bibitem[Draine et al.(2007)]{dra07} Draine, B.T., Dale, D.A., Bendo, G., Gordon, K.D., Smith, J.D.T., 
Armus, L., Engelbracht, C.W., Helou, G., Kennicutt, R.C., Li, A., et al.  2007, ApJ, accepted 
(astroph/073213)
\bibitem[Eales et al.(1999)]{eale99} Eales, S., Lilly, S., Gear, W., Dunne, L., Bond, J.R., Hammer, F., 
Le Fevre, O., \& Crampton, D. 1999, \apj, 515, 518
\bibitem[Engelbracht et al.(2005)]{enge05} Engelbracht, C.W., Gordon, K.D., Rieke, G.H., Werner, 
M.W., Dale, D.A., \& Latter, W.B.  2005, ApJ, 628, 29
\bibitem[F\"orster Schreiber et al.(2004)]{forst04} F\"orster Schreiber, N.M.,
  Roussel, H., Sauvage, M., \& Charmandaris, V., 2004, \aap, 419, 501
 \bibitem[Gallagher, Hunter \& Bushouse(1989)]{gal89} Gallagher, J.S., Hunter, D.A., \& Bushouse, 
 H. 1989, \aj, 97, 700
\bibitem[Giavalisco et al.(2004)]{gia04} Giavalisco, M., et al. 2004, \apj, 600, L103
\bibitem[Gordon et al.(2004)]{gor04} Gordon, K.D., Perez--Gonzalez,
P.G., Misselt, K.A., Murphy, E.J., Bendo, G.J., Walfer, F., Thornely,
M.D., Kennicutt, R.C., et al. 2004, \apjs 154, 215 
\bibitem[Haas, Klaas \& Bianchi(2002)]{haas02} Haas, M., Klaas, U., \&
Bianchi, S. 2002, \aap, 385, L23
\bibitem[Heckman et al.(1998)]{heck98} Heckman, T.M., Robert, C., Leitherer,
  C., Garnett, D.R., \& van der Rydt, F. 1998, \apj, 503, 646
\bibitem[Helou(1986)]{hel86} Helou, G. 1986, \apj, 311, L33 
  \bibitem[Helou et al.(2004)]{helo04} Helou, G., Roussel, H., Appleton, P.,
Frayer, D., Stolovy, S., Storrie--Lombardi, L., Hurst, R., Lowrance, P., et
al. 2004, \apjs, 154, 253
\bibitem[Hopkins(2004)]{hop04} Hopkins, A.M. 2004, \apj, 615, 209
\bibitem[Hopkins \& Beacom(2006)]{hop06} Hopkins, A.M., \& Beacom, J.F. 2006, \apj, 651, 142
\bibitem[Hopkins et al.(2001)]{hop01} Hopkins, A. M., Connolly, A. J., Haarsma, D. B., \& Cram, 
L. E. 2001, \aj, 122, 288
\bibitem[Hughes et al.(1998)]{hug98} Hughes, D.H., Serjeant, S., Dunlop, J., Rowan-Robinson, M., 
Blain, A., Mann, R.G., Ivison, R. Peacock, J., Efstathiou, A., Gear, W., et al. 1998, Nature, 394, 241
\bibitem[Hunter et al.(1986)]{hun86} Hunter, D.A., Gillett, F.C., Gallagher, J.S., Rice, W.L., \& Low, 
F.J. 1986, \apj, 303, 171
\bibitem[Iglesias--Paramo et al.(2006)]{igl06} Iglesias--Paramo, J., et al. 2006, \apjs, 164, 38
\bibitem[Jansen, Franx \& Fabricant(2001)]{jan01} Jansen, R.A., Franx, M., \& Fabricant, D. 2001, 
\apj, 551, 825
\bibitem[Katz(1992)]{kat92} Katz, N. 1992, \apj, 391, 502
\bibitem[Kennicutt(1998a)]{ken98a} Kennicutt, R.C. 1998a, \apj, 498, 541
\bibitem[Kennicutt(1998b)]{ken98b} Kennicutt, R.C. 1998b, \araa, 36, 189
\bibitem[Kennicutt et al.(2003)]{ken03} Kennicutt, R.C., Armus, L., Bendo,
  G., Calzetti, D., Dale, D.A., Draine, B.T., Engelbracht, C.W., Gordon, D.A.,
  et al. 2003a, \pasp, 115, 928
\bibitem[Kennicutt et al.(2007)]{ken07} Kennicutt, R.C., Calzetti, D., et al. 2007, \apj, submitted
\bibitem[Kewley et al.(2002)]{kew02} Kewley, L.J., Geller, M.J., Jansen, R.A., \& Dopita, M.A. 2002, \aj, 124, 3135
\bibitem[Kewley, Geller \& Jansen(2004)]{kew04} Kewley, L.J., Geller, M.J., \& Jansen, R.A. 2004, 
\aj, 127, 2002
\bibitem[Kong et al.(2004)]{kon04} Kong, X., Charlot, S., Brinchmann,
J., \& Fall, S.M. 2004, \mnras, 349, 769
\bibitem[Leger \& Puget(1984)]{lege84} Leger, A., \& Puget, J.L. 1984,
\aap, 137, L5
\bibitem[Leitherer et al.(1999)]{lei99} Leitherer, C., Schaerer, D., 
Goldader, J.D., Gonz\'alez Delgado, R.M., Robert, C., Kune, D.F., de Mello, 
D.F., Devost, D., \& Heckman, T.M. 1999, \apjs, 123, 3
\bibitem[Li \& Draine(2002)]{li02} Li, A., \& Draine, B.T. 2002, \apj, 572, 762 
\bibitem[Lilly et al.(1996)]{lil96} Lilly, S. J., Le Fvre, O., Hammer, F., \& Crampton, D. 1996, \apj, 460, L1
\bibitem[Lonsdale Persson \& Helou(1987)]{per87} Lonsdale Persson, C.J., \& Helou, G.X. 
1987, \apj, 314, 513
\bibitem[Madden et al.(2006)]{madd06}  Madden, S.C., Galliano, F.,  Jones, A.P., \& 
Sauvage, M.  2006, \aap,  446, 877
\bibitem[Madau et al.(1996)]{mad96} Madau, P., Ferguson, H.C., Dickinson, M.E., Giavalisco, M., 
Steidel, C.C., \& Fruchter, A. 1996, \mnras, 283, 1388
\bibitem[Madau, Pozzetti \& Dickinson(1998)]{mad98} Madau, P., Pozzetti, L., \& Dickinson, M.E. 
1998, \apj, 498, 106
\bibitem[Martin et al.(2005)]{mar05} Martin, D.C., Fanson, J., Schiminovich, D., Morrissey, P., 
Friedman, P.G., Barlow, T.A., Conrow, T., Grange, R., Jelinsky, P.N., et al. 2005, \apj, 619, L1
\bibitem[Mattioda et al. (2005)]{matt05} Mattioda, A.L., Allamandola, L.J., \& Hudgins, D.M. 
 2005, \apj, 629, 1183
\bibitem[Meurer, Heckman \& Calzetti(1999)]{meu99} Meurer, G.R., Heckman,
  T.M., \& Calzetti, D. 1999, \apj, 521, 64 
 \bibitem[Moustakas, Kennicutt \& Tremonti(2006)]{mou06} Moustakas, J., Kennicutt, R.C., \& 
Tremonti, C.A. 2006, ApJ, 642, 775
\bibitem[Norman et al.(2004)]{nor04} Norman, C., et al. 2004, \apj, 607, 721
\bibitem[Peeters, Spoon \& Tielens(2004)]{peete04} Peeters, E., Spoon,
H.W.W., \& Tielens, A.G.G.M. 2004, \apj, 613, 986
\bibitem[Perez--Gonzalez et al.(2006)]{gonz06} Perez--Gonzalez, P.G., Kennicutt, R.C., Gordon, K.D., 
Misselt, K.A., Gil de Paz, A., Engelbracht, C.W., Rieke, G.H., Bendo, G.J., Bianchi, L., Boissier, S., 
Calzetti, D., Dale, D.A., et al.  2006, \apj, 648, 987
\bibitem[Pety et al.(2005)]{pety05} Pety, J., Teyssier, D., Fosse`,
D., Gerin, M., Roueff, E., Abergel, A., Habart, E., \& Cernicharo,
J. 2005, \aap, in press (astroph/0501339)
\bibitem[Ranalli, Comastri \& Setti(2003)]{ran03} Ranalli, P., Comastri, A., \& Setti, G. 2003, \aap, 399, 
39
\bibitem[Rosa--Gonzalez, Terlevich \& Terlevich(2002)]{ros02}
Rosa--Gonzalez, D., Terlevich, E., \& Terlevich, R. 2002, \mnras, 332,
283
\bibitem[Roussel et al.(2001)]{rous01} Roussel, H., Sauvage, M.,
Vigroux, L., \& Bosma, A. 2001, \aap, 372, 427
\bibitem[Rowan--Robinson \& Crawford(1989)]{row89} Rowan--Robinson, M., \& Crawford, 
J. 1989, \mnras, 238, 523
\bibitem[Salim et al.(2007)]{sal07} Salim, S., Rich, M.R., Charlot, S., Brinchmann, J., Johnson, B.D., 
Schminovich, D., Seibert, M., Mallery, R., Heckman, T.M., Forster, K., et al. 2007, \apjs, in press 
(astroph/0704.3611)
\bibitem[Sauvage \& Thuan(1992)]{sau92} Sauvage, M., \& Thuan, T.X. 1992, \apj, 396, L69
\bibitem[Schmitt et al.(2006)]{sch06} Schmitt, H.R., Calzetti, D., Armus, L., Giavalisco, M., Heckman, 
T.M., Kennicutt, R.C., Leitherer, C., \& Meurer, G.R. 2006, \apj, 643, 173
\bibitem[Seibert et al.(2005)]{sei05} Seibert, M. et al. 2005, \apj, 619, L55
\bibitem[Sellgren(1984)]{sell84} Sellgren, K. 1984, \apj, 277, 623
\bibitem[Smail, Ivison \& Blain(1997)]{sma97} Smail, I., Ivison, R.J., \& Blain, A.W. 1997, \apj, 490, L5
\bibitem[Springel \& Hernquist(2003)]{spr03} Springel, V., \& Hernquist, L. 2003, \mnras, 339, 289
\bibitem[Springel, di Matteo  \& Hernquist(2005)]{spr05} Springel, V., di Matteo, T. \& Hernquist, L. 2005, \mnras, 361, 776
\bibitem[Steidel et al.(1999)]{ste99} Steidel, C.C., Adelberger, K.L., Giavalisco, M., Dickinson, M., 
\& Pettini, M. 1999, \apj, 519, 1
\bibitem[Stinson et al.(2006)]{sti06} Stiinson, G., Seth, A., Katz, N., Wadsley, J., Governato, F., 
Quinn, T. 2006, \mnras, 373, 1074
\bibitem[Sullivan et al.(2001)]{sul01} Sullivan, M., Mobasher, B., Chan, B., Cram, L., Ellis, R., 
Treyer, M., \& Hopkins, A. 2001, \apj, 558, 72
\bibitem[Wang \& Heckman(1996)]{wan96} Wang, B., \& Heckman, T.M. 1996, \apj,
  457, 645
\bibitem[Wilson et al.(2002)]{wil02} Wilson, G., Cowie, L.L., Barger, A.J., \& Burke, D.J. 2002, \aj, 124, 
1258
\bibitem[Wu et al.(2005)]{wu05}  Wu, H., Cao, C., Hao, C.-N., Liu, F.-S., 
Wang, J.-L., Xia, X.-Y., Deng, Z.-G., \& Young, C. K.-S. 2005, \apj, 632, L79 
\bibitem[Yan et al.(1999)]{yan99} Yan, L., McCarthy, P.J., Freudling, W., Teplitz, H.I., Malumuth, E.M., 
Weymann, R.J., \& Malkan, M.A. 1999, \apj, 519, L47
\end{thebibliography}
\end{document}